\documentclass[runningheads,a4paper]{llncs}

\usepackage{amssymb}
\usepackage{graphicx}
\usepackage{verbatim}
\usepackage{url}
\urldef{\mailsa}\path|rastislav.lenhardt@comlab.ox.ac.uk|
\urldef{\mailsb}\path||
\urldef{\mailsc}\path|}|

\begin{document}

\mainmatter  

\title{Proof of Concept: \\ Fast Solutions to NP-problems by Using SAT and Integer Programming Solvers }

\titlerunning{Fast Solutions to NP-problems by Using SAT and ILP}

\author{Rastislav Lenhardt}

\institute{Computing Laboratory, University of Oxford, United Kingdom \\ \mailsa\\}

\maketitle

\section{Introduction}

In the last decade, the power of the state-of-the-art SAT and ILP (Integer Linear Programming) solvers has dramatically increased. They implement many new techniques and heuristics and since any NP problem can be converted to SAT or ILP instance, we could take advantage of these techniques in general by converting the instance of NP problem to SAT formula or Integer program. 

Cook-Levin theorem says that any NP problem can be reduced to SAT of polynomial size. However, to maximize the speed of our algorithms, we would like to get as short instance as possible (and with as least number of variables as possible). This concept of SAT complexity alone is worth more research in future.

A problem we consider, in this proof of concept, is $k$-Clique problem.

\begin{definition}
Let $G = (V,E)$ be an undirected graph on $n$ vertices $V$ with edges $E \subset V \times V$. Then $k$-Clique decision problem asks if there exists a set $S \subset V$ such that $|S|=k$ and for any two vertices $x, y$ we have $x \in S \wedge y\in S \rightarrow (x,y) \in E$.
\end{definition}

\section{Fast Backtrack Algorithm}

Naive approach to $k$-Clique problem would be to enumerate all subsets of $V$ of size $k$ and then check if all vertices among the subset are connected. It would take at least ${n \choose k}$ steps. For example for $n=100$ and $k=23$ it is approximately $2.486 \times 10^{22}$ and so on the standard personal computer it would run at least $300$ thousand years.

Much better approach in practice is to traverse only cliques of size up to $k$ in the graph, instead of traversing all subsets of size $k$.

To do so, we start with a clique of size one and then try to add next vertex at a time, such that it is connected to all the previous ones that are already in the clique. If we cannot add any more, we remove the last one, and try to add verticies with the higher index. We terminate if we find a clique of size $k$. In this approach, we traverse only cliques in an input graph which are also of size up to $k$, so we can get much better performance than in naive approach. However, as we see in the results of experiments later, even this backtrack specially tailored for this problem, is very slow on some instances when we compare it to approaches from the next sections. An implementation of backtrack used in our experiments follows:

\begin{verbatim}
#include<iostream>
#include<vector>
using namespace std;
#define VI vector<int>
#define PB push_back

int N,K;

int main() {
    cin>>N>>K;
    vector<VI> a = vector<VI>(N,VI(N,0));
    // read input
    for (int i=0;i<N;i++) for (int j=0;j<N;j++) cin>>a[i][j];
    
    VI b = VI(K,-1);
    b[0]=0;
    int index=1;
    while (index>=0&&index<K) {
          bool found=false;
          int start = b[index]+1;
          if (index>0&&b[index-1]>start) start=b[index-1]+1; 
          for (int t=start;t<N;t++) {
              bool ok=true;
              for (int k=0;k<index;k++) 
                  if (a[b[k]][t]==0) { ok=false; break;}
              if (ok) { 
                      b[index]=t; index++; 
                      for (int k=index;k<N;k++) b[k]=-1; 
                      found = true; 
                      break;
                      }    
          }
          if (!found) {
             index--;            
          }
    } 
    if (index==-1) cout<<"UNSATASFIABLE"<<endl;
    if (index==K) { 
                  cout<<"SATISFIABLE"<<endl;
                  for (int i=0;i<b.size();i++) cout<<b[i]<<" ";cout<<endl;
    }
}
\end{verbatim}

\section{Translation to SAT}

An input is an integer $k$ and a binary matrix $A_{n \times n}$, where $A[i,j] = 1$ iff vertices $i$ and $j$ are adjacent.

We introduce literalls $x_i$ for each vertex of $V = \{ 1, 2, \ldots n \}$. The value of $x_i$ is true (or $1$) if we choose vertex $i$ to be part of the $k$-clique and  the value is false (or $0$) otherwise.

We build formula $\phi_1$ to ensure the required behaviour, i.e. that if we choose vertices $i$ and $j$ to our clique then there must be an edge $(i,j)$:

$$\phi_1 = \bigwedge_{i,j} (x_i \wedge x_j \rightarrow A[i][j])$$

It can be easily translated to conjunctive normal form (CNF):

$$\phi_1 = \bigwedge_{i,j} (\neg x_i \vee \neg x_j \vee  A[i][j])$$

The next property we need to ensure is that exactly $k$ literalls $x_i$ are true. We introduce unary counters to do so (experiments in encoding some problems to SAT showed that current SAT solvers perform as well with unary as with binary counters despite the fact that we introduce more variables). 

Each unary counter will be set of $n+1$ literalls $c_0, c_1, \ldots c_k$ such that exactly one of them is true. Then this truth literall deonotes the value of counter (between $0$ and $k$, inclusive). First we have counter $c^0$ set to $0$ at the beginning and then we introduce counter $c^i$ for each $x_i$ such that by abuse of notation $c^{i+1} = c^i + x_i$. For example, if counter $3$ has value $7$ then $c^3_7$ is true.

Formula $\phi_2$ ensures that each counter has exactly only one set value:
$$\phi_2 = \bigwedge_k ( (\bigwedge_{i \neq j} c^k_i \rightarrow \neg c^k_j)
\vee (\bigvee_i c^k_i))$$

Formula $\phi_3$ ensures that counter is initially set to $0$:
$$\phi_3 = c^0_1 \wedge \bigwedge_{i>1} \neg c^0_i $$

Formula $\phi_4$ ensures that counters values update well if $x_i$ is false:
$$\phi_4 = \bigwedge_i \neg x_i \rightarrow \bigwedge_j c^i_j \leftrightarrow c^{i-1}_j$$

Formula $\phi_5$ ensures that counters values update well if $x_i$ is true:
$$\phi_5 = \bigwedge_i x_i \rightarrow \bigwedge_{j>0} c^i_j \leftrightarrow c^{i-1}_{j-1}$$

Finally, we get one large formula $\phi = \phi_1 \wedge \phi_2 \wedge \phi_3 \wedge \phi_4 \wedge \phi_5$, which can be transformed to CNF.

\subsection{An Input Format Used by SAT Solvers}

We need to transform $\phi$ to CNF, because that is the standard input format used by current SAT-solvers. It is also the standard used by annual SAT solvers competitions. You can find its whole definition in \cite{format}. 

The format maps each variable to a positive integer if it occurs in a positive form (i.e. to $i$ for $x_i$) or to corresponding negative integer if it occurs in a negative form (i.e. to $-i$ for $\neg x_i$). Then all clauses are split by $0$. For example, formula $(x_1 \vee \neg x_2) \wedge (\neg x_4 \vee \neg x_1 \vee x_3)$ is transformed to \begin{verbatim} 1 -2 0 -4 -1 3 0 \end{verbatim}

\subsection{Implementation}

\begin{verbatim}
#include<iostream>
#include<vector>
using namespace std;
#define VI vector<int>
#define PB push_back

int N,K;
int vars;

void c(int i, int j, bool t) {
     int x = N+i*(K+1)+j+1;
     if (!t) x*=-1;
     cout<<x<<" ";
}

void x(int i, bool t) {
     if (!t) i*=-1;
     cout<<i<<" ";     
}

int clauses;
void cl_end() {
     cout<<"0"<<endl;     
     clauses++;
}

void intro() {
     cout<<"p cnf "<<vars<<" "<<clauses<<endl;
}

int main() {
    cin>>N>>K;
    vars = N+(N+1)*(K+1);
    clauses=0;
    vector<VI> a = vector<VI>(N,VI(N,0));
    // read input
    for (int i=0;i<N;i++) for (int j=0;j<N;j++) cin>>a[i][j];
    // set counters
    for (int i=0;i<=K;i++) { if (i==0) c(0,i,true); else c(0,i,false); cl_end();}
    for (int i=1;i<=N;i++) {
        c(i,1,false); c(i,0,false); cl_end();    
    }
    c(N,K,true); cl_end();

    // edge satisfiability
    for (int i=0;i<N;i++) for (int j=i+1;j<N;j++) {
        if (a[i][j]==0) {
           x(i+1,false); x(j+1,false); cl_end();                
        }
    }

    // counter update if (x_i ==0 )
    for (int i=1;i<=N;i++) for (int j=0;j<=K;j++) {
        c(i-1,j,false); c(i,j,true); x(i,true); cl_end();
        c(i,j,false); c(i-1,j,true); x(i,true); cl_end();
    }    
    // counter update if (x_i == 1)
    for (int i=1;i<=N;i++) for (int j=1;j<=K;j++) {
        c(i-1,j-1,false); c(i,j,true); x(i,false); cl_end();
        c(i,j,false); c(i-1,j-1,true); x(i,false); cl_end();
    }
    
    intro(); 
        
}
\end{verbatim}

\section{Translation to Integer Linear Programming}

We use the similar idea as in translation to SAT. Now we have binary variables $x_i$ for each vertex and we assign them value $1$ if and only if vertex $i$ is in the clique. We will translate input integer $k$ and matrix $A_{n \times n}$ representing graph to an integer program that will not only decide if there is a clique of size $k$, but we will maximize the size of this clique.

So we want to compute $$\mathop{max} \sum_i x_i$$ subject to conditions that $x_i=1 \wedge x_j=1 \rightarrow A[i][j]$. These conditions can be written as linear inequalities 
$$ x_i + x_j \le 1+A[i][j] \mathrm{~~~} (\forall i, j)$$

The listing of translation programme:

\begin{verbatim}
#include<iostream>
#include<vector>
using namespace std;
#define VI vector<int>
#define PB push_back
int N,K;
int vars;

int main() {
    cin>>N>>K;
    vector<VI> a = vector<VI>(N,VI(N,0));
    // read input
    for (int i=0;i<N;i++) for (int j=0;j<N;j++) cin>>a[i][j];
    
    for (int i=0;i<N;i++) { 
        if (i!=0) cout<<"+";
        cout<<"x"<<i; 
    }
    cout<<";"<<endl;
    //cout<<" = "<<K<<";"<<endl;
    for (int i=0;i<N;i++) {
        cout<<"0 <= x"<<i<<" <=1;"<<endl;
    }
    for (int i=0;i<N;i++) for (int j=i+1;j<N;j++) {
        cout<<"x"<<i<<" + x"<<j<<" <= "<<1+a[i][j]<<";"<<endl;    
    }
        cout<<"int ";
    for (int i=0;i<N;i++) {
        if (i!=0) cout<<",";
        cout<<"x"<<i;
    }
    cout<<";"<<endl;
    
     
}
\end{verbatim}

\section{Generating random instances}

We have built generator of random instances of graphs to be able to execute comparison tests. Of course, performance could have been different if we chose another way of generating instances for the performance tests. However, the results on random instances shall give a good indication, because they are usually more difficult than real world instances.

We generate graph on $n$ vertices such that probability of each edge being in graph is uniformly and independently $A/B$, where $n$, $A$ and $B$ are provided as input.

\begin{verbatim}
#include<iostream>
#include<vector>
using namespace std;
#define VI vector<int>
#define PB push_back

int main() {
    int N,K;
    int A,B;
    srand(time(0));
    cin>>N>>K>>A>>B;

    // A/B is probability
    vector<VI> a = vector<VI>(N,VI(N,0));
    for (int i=0;i<N;i++) for (int j=i+1;j<N;j++) {
        if (rand()%B<A) a[i][j]=1; a[j][i]=a[i][j];    
    }    
    cout<<N<<" "<<K<<endl;
    for (int i=0;i<N;i++) {
        for (int j=0;j<N;j++) cout<<" "<<a[i][j];
        cout<<endl;   
    }
}

\end{verbatim}

\section{Experimental results}

The main purpose of this report is to compare three different approaches to solve the instance of $k$-Clique problem.

\begin{enumerate}
\item Optimized backtrack solution.
\item Translate problem to SAT and use SAT solver. We chose MiniSAT (and its Java implementation Sat4j \cite{sat4j}) as one of the best performing state-of-the-art SAT solvers.
\item Translate problem to Integer Linear Programming and use ILP solver. We chose lp\_solve 5.5 \cite{lpsolver} as one of the best current ILP solvers.
\end{enumerate}

We refer to the instances solved by the approaces above as to back$(n,k,a/b)$, sat$(n,k,a/b)$ and ilp$(n,a/b)$, where $n$ is a number of vertices of graph, $k$ is a size of clique we look for and $a/b$ is an independent probability for any edge of the random graph being present.

We ran several tests on Intel Core2 Duo CPU @2.10 Ghz with 4 GB of RAM and found several interesting results. Once we generated random graph, we used the same graph for different values of $k$ to test all approaches.

\subsection{Smaller Instances}
First we considered smaller instances, where resulting largest clique is of size less than $14$. We look at graphs on $40$, $60$ an $100$ vertices with edge probability being $2/3$.

This table shows the performance of sat$(40,*,2/3)$, sat$(60,*,2/3)$ and also how efficient the translation is:

\begin{table}[ht]
\centering
\caption{Graphs on $40$ and $60$ vertices; SAT solver}
\begin{tabular}{c|c|c|c|c}
instance & \# variables & \# clauses & time & satisfiability \\
\hline
sat$(40,8,2/3)$ & $409$ & $1639$ &  $0.088$ s & SAT \\
sat$(40,10,2/3)$ & $491$ & $1957$ &  $0.318$ s & SAT \\
sat$(40,11,2/3)$ & $521$ & $2116$ &  $0.381$ s & UNSAT \\
sat$(40,12,2/3)$ & $573$ & $2275$ &  $0.384$ s & UNSAT \\
\hline
sat$(60,11,2/3)$ & $792$ & $3377$ &  $0.050$ s & SAT \\
sat$(60,12,2/3)$ & $853$ & $3616$ &  $1.629$ s & SAT \\
sat$(60,13,2/3)$ & $914$ & $3855$ &  $2.891$ s & UNSAT \\
sat$(60,14,2/3)$ & $975$ & $4094$ &  $2.622$ s & UNSAT \\
sat$(60,16,2/3)$ & $1097$ & $4572$ &  $2.391$ s & UNSAT \\
\end{tabular}
\end{table}

and we can compare it with competing approaches:

\begin{table}[ht]
\centering
\caption{Graphs on $40$ and $60$ vertices; ILP solver and backtrack}
\begin{tabular}{c|c|c}
instance & time & comment \\
\hline
ilp$(40,2/3)$ & $0.185$ s & result is $10$ \\
back$(40,8,2/3)$ & $0.102$ s &  SAT \\
back$(40,10,2/3)$ & $0.212$ s & SAT \\
back$(40,11,2/3)$ & $0.277$ s & UNSAT \\
back$(40,12,2/3)$ & $0.260$ s & UNSAT \\

\hline
ilp$(60,2/3)$ & $1.440$ s & result is $12$ \\
back$(60,11,2/3)$ & $0.171$ s & SAT \\
back$(60,12,2/3)$ & $0.177$ s & SAT \\
back$(60,13,2/3)$ & $2.175$ s & UNSAT \\
back$(60,14,2/3)$ & $2.178$ s & UNSAT \\
back$(60,16,2/3)$ & $2.225$ s & UNSAT \\
\end{tabular}
\end{table}

\begin{table}[ht]
\centering
\caption{Graph on $100$ vertices, where the largest clique is of size $13$}
\begin{tabular}{c|c|c}
instance & time & comment \\
\hline
ilp$(100,2/3)$ & $117$ s & result is $13$ \\
sat$(100,12,2/3)$ & $7.5$ s& SAT \\
sat$(100,13,2/3)$ & $16$ s& SAT \\
sat$(100,14,2/3)$ & $152$ s& UNSAT \\
back$(100,12,2/3)$ & $0.5$ s& SAT \\
back$(100,13,2/3)$ & $6$ s& SAT \\
back$(100,14,2/3)$ & $60$ s& UNSAT \\
\end{tabular}
\end{table}

\subsection{Larger Instances}

We tried not only one, but several instances of random graph on $100$ vertices with edge probability $5/6$. Times were quite similar from the viewpoint of conclusions we can make out of them. This concrete problem was especially interested also from the point of view of Random Graphs. 

One can try to predict the size of the largest clique in the random graph with these parameters (answering the question, what is the most probable result). It is not simple task. It is possible to get some idea by solving the question what is the expected number of cliques of size $k$ in such graph for different $k$. In this case, it might suggest that the size of the largest clique should be about $24$ or $25$, strictly because the expected number of cliques of size $24$ in a random graph $(100, 5/6)$ is over a hundred. However the experiments did not confirm it, with the largest clique being most often $22$ or $21$. The explanation might be that it is quite rare that we have larger clique, but once we have one, it inducess also huge number of smaller ones.

\begin{table}[ht]
\centering
\caption{Graph on $100$ vertices, where the largest clique is of size $22$}
\begin{tabular}{c|c|c}
instance & time & comment \\
\hline
ilp$(100,5/6)$ & $417$ s & result is $22$ \\
sat$(100,22,5/6)$ & $900$ s& SAT \\
sat$(100,23,5/6)$ & $20\mathrm{~}000$ s& UNSAT \\
back$(100,22,5/6)$ & $>50\mathrm{~}000$ s& timeout \\
back$(100,23,5/6)$ & $>50\mathrm{~}000$ s& timeout \\

\end{tabular}
\end{table}

\newpage

\section{Interpretation of Experimental Results}
The interpretation of above results suggests that ILP is the best method for solving $k$-Clique problem. It scales very well and we can see its speed especially for large instances, when the size of largest clique is approaching $20$ and above. Its speed is very stable for different random instances with the same parameters and it can usually find the optimal solution very quickly. It spends most of the time on proving that there is no better solution.

It is similar for SAT method, which works the slowest for $k$ being the first unsatisfiable or the last satisfiable. Its performance gets better when it tries to prove that there is not a clique of size $k$ for $k$ getting furher away from the size of largest clique. Despite the fact that SAT seems slower than ILP for this particular problem, it still outperforms backtrack tremendously on large instances. And it is without the necessity of being clever and thinking of the best backtrack for this problem.

On the other hand, this tailored backtrack is very fast on small instances, where it is easy to find a solution. One of the reason is that it does not have any starting costs comparing to competitors. For larger SAT formulas or ILPs, preprocessing phase takes non-trivial time, but it is trade-off for much better performances on large instances later.

\section{Future Work}

It would be intersting to make the similar study about NP problem that is less quantitative. For example finding some colouring of the graph. It may happen that for such problems SAT would be better than ILP. 

We do not answer the question if SAT or ILP approch is better in general. However, we can confirm that improvements in techniques of solving SAT and ILP problems from the last decade can be very beneficial to solve other NP problems. Not only it allows us not to think about tailored backtracks case by case for different NP problems, but its performance completely outperforms the tailored backtrack for large instances. 

Therefore, as future work, we suggest to look for the most effective translations of other NP problems to either SAT or ILP problems.


\begin{thebibliography}{4}

\bibitem{sat4j}
Sat4j solver:
http://www.sat4j.org/

\bibitem{lpsolver} 
LP solve 5.5: 
http://lpsolve.sourceforge.net/5.5/

\bibitem{satcontests} 
Annual SAT competitions:
http://www.satcompetition.org/

\bibitem{format} 
Common CNF Dimacs format:
http://www.cs.ubc.ca/\~ {}hoos/ \\
SATLIB/Benchmarks/SAT/satformat.ps
\end{thebibliography}
\end{document}